\documentclass[aps, prl, reprint, showpacs, twocolumn, 10pt, groupedaddress]{revtex4-1}

\usepackage{graphicx, amsmath, dsfont, dcolumn, bm, times, color, braket, amssymb, multirow}

\begin{document}

\bibliographystyle{apsrev4-2}

\title{Current-voltage characteristics of the N-I-PT-symmetric non-Hermitian superconductor junction as a probe of non-Hermitian formalisms}

\author{Viktoriia Kornich}
\affiliation{Institute for Theoretical Physics and Astrophysics, University of W\"urzburg, 97074 W\"urzburg, Germany }

\date{\today}

\begin{abstract}
We study theoretically a junction consisting of a normal metal, PT-symmetric non-Hermitian superconductor, and an insulating thin layer between them (N-I-PTS junction). We calculate current-voltage characteristics for this junction using left-right and right-right bases and compare the results. We find that in the left-right basis, the Andreev-scattered particles move in the opposite direction compared to the right-right basis and conventional Andreev scattering. This leads to profound differences in current-voltage characteristics. Based on this and other signatures, we argue that left-right basis is not applicable in this case. Remarkably, we find that the growth and decay with time of the states with imaginary energies in right-right basis are equilibrated.
\end{abstract}

\maketitle

\let\oldvec\vec
\renewcommand{\vec}[1]{\ensuremath{\boldsymbol{#1}}}
{\it Introduction.} 
Non-Hermitian systems are of high interest due to numerous potential applications, exotic behaviour \cite{ashida:advphys21, bergholtz:rmp21}, and even reformulated quantum mechanics formalism \cite{moiseyev:book11, meden:arxiv23}. Among them, PT-symmetric non-Hermitian systems \cite{bender:prl98} have brought new effects in optics and photonics \cite{feng:natph17, elganainy:natphys18, ozdemir:natmater19, guo:prl09, regensburger:nature12}. PT-symmetry denotes combined space-inversion symmetry (parity) $\mathcal{P}$ and time-inversion symmetry $\mathcal{T}$. In condensed matter physics, non-Hermitian formalism and in particular PT-symmetric systems have also started to receive interest \cite{ghatak:prb18, kawabata:prx19, zhou:prb20, kawabata:prb18, okugawa:prb19, klett:pra17, wang:prb21, xian:arxiv23}. This topic is developing and many chapters characteristic for condensed matter physics are still missing. For example, transport and thermodynamical properties of non-Hermitian systems, where imaginary energies gives divergences \cite{arouca:qf22}, are largely not understood. Among known related works are the ones about wave transport in non-Hermitian tight-binding models \cite{eichelkraut:natcom13, ghaemi:pra21, zhang:prb19, longhi:prb17, hoeckendorf:aplph21} and semiclassical equations of motion for Bloch electrons \cite{xu:prl17, silberstein:prb20}. 

Non-Hermitian Hamiltonians $H$ in general have non-orthogonal eigenstates $H|R_j\rangle=E_j|R_j\rangle$, $\langle R_j|R_i\rangle\neq \delta_{i,j}$ \cite{ashida:advphys21, bender:rpp07, mostafazadeh:jmp02} forming so-called right-right (RR) basis.  This non-orthogonality can be corrected by introducing a different scalar product, e.g. $\mathcal{CPT}$-product \cite{bender:rpp07}, $\eta$-product for pseudo-Hermitian systems \cite{mostafazadeh:jmp02}, or left-right (LR) basis with the left eigenstate defined as $H^\dagger|L_j\rangle=E_j^*|L_j\rangle$ and $\langle L_j|R_i\rangle=\delta_{i,j}$. Both RR and LR bases are used in modern physics \cite{herviou:scipost19}, however they can give different results. For example, the probability of the eigenstate with imaginary energy grows or decays with time in RR basis, while it is constant in LR basis \cite{herviou:scipost19}. We think that in order to understand, which basis to use and what are the limits of their applicability, we need to consider a physical observable and compare it in LR and RR bases. 

In this work, we study current-voltage characteristics of a junction formed by a normal metal, an insulating thin layer, and a PT-symmetric non-Hermitian superconductor (N-I-PTS junction) in LR and RR bases and compare them. We show that LR basis gives drastically different Andreev scattering: the Andreev-scattered particles move into the direction of the junction, not away from it. Thus, in LR basis Andreev scattering gives negative shift of current in contrast to RR case and conventional N-I-S junctions \cite{blonder:prb81}. We analyze the bands with imaginary energies in LR and RR formalisms and in particular the distribution function there, that formally has divergences at imaginary energies. We argue that $\Re{[E]}$ should be used in the distribution function at least in LR basis. Remarkably, the decaying and growing in time states in RR basis appear to be connected through the source term and thus the overall probability is conserved instead of exponential growth that is often obtained for PT-symmetric systems.

{\it PT-symmetric non-Hermitian superconductor.}
The Hamiltonian of PTS in the basis $\Psi^\dagger(x)=(\psi^\dagger_\uparrow(x)  \ \ \psi_\downarrow(x))$ is
\begin{eqnarray}
\mathcal{H}_{\rm PTS}=\int \Psi^\dagger(x)\begin{pmatrix}-\frac{\partial_x^2}{2m}-\mu & -i\Delta\partial_x\\ i\Delta\partial_x & \frac{\partial_x^2}{2m}+\mu  \end{pmatrix}\Psi(x)dx,
\end{eqnarray}
where $m$ is the effective mass of an electron, $\mu$ is the chemical potential, and $\psi_\sigma(x)$ is an electron field annihilation operator with the spin $\sigma$. The superconducting part with the mean field $\Delta$ has $p$-wave symmetry, but is non-Hermitian. In the momentum representation, in the basis $\Psi^\dagger(k)=( \psi^\dagger_\uparrow(k) \ \  \psi_\downarrow(-k))$, it is
\begin{eqnarray}
H_{\rm PTS}(k)=\begin{pmatrix}\frac{k^2}{2m}-\mu & \Delta k\\ -\Delta k & -\frac{k^2}{2m}+\mu  \end{pmatrix}.
\end{eqnarray}
It is PT-symmetric, $(\mathcal{PT})H_{\rm PTS}(k)(\mathcal{PT})^{-1}=H_{\rm PTS}(k)$.

Physically, the opposite signs of the off-diagonal superconducting terms in $H_{\rm PTS}(k)$ implies that electron-electron interaction is asymmetric: $H_{\rm e-e}=\sum_{p,q}\psi^\dagger_{\uparrow}(p+q)\psi^\dagger_\downarrow(-p-q)W(q)\psi_\downarrow(-p)\psi_\uparrow(p)$ and $W(q)=-W(-q)$. Thus, when electrons interact attractively, the corresponding holes interact repulsively and vice versa. We have proposed that such interaction can occur due to the spatiotemporal modulations of the material \cite{kornich:prr22a}, which can induce assymmetry of the phonon spectrum and consequently asymmetric phonon-mediated electron-electron interaction.
 
 The spectrum of PTS is $E_k=\pm\sqrt{\left(\frac{k^2}{2m}-\mu\right)^2-(\Delta k)^2}$. Thus, there is no gap, in contrast to conventional superconductors, and there is a regime of imaginary energies $E_k$ (see Fig. \ref{fig:NIPTSelectronscattering}, right). This regime is called PT-broken regime, while the regime with real $E_k$ is PT-unbroken \cite{bender:rpp07}. These two regimes are connected by the exceptional points at $E=0$. 
 
{\it Charge continuity equation.}
We now consider charge evolution in PTS, where the charge operator is $Q(x)=e\rho(x)=e\sum_{\sigma=\uparrow,\downarrow}\psi^\dagger_\sigma(x)\psi_\sigma(x)$. We can define its average using LR basis or RR basis, yielding different evolution equations \cite{herviou:scipost19,supplmaterial}. In LR basis, it is the usual Heisenberg evolution equation:
\begin{eqnarray}
\label{eq:EvEqLR}
\frac{d}{dt}Q(x)=i[\mathcal{H}_{\rm PTS}, Q(x)],
\end{eqnarray}
while in RR basis it is
\begin{eqnarray}
\label{eq:EvEqRR}
\frac{d}{dt}Q(x)=i(\mathcal{H}_{\rm PTS}^\dagger Q(x)-Q(x)\mathcal{H}_{\rm PTS}).
\end{eqnarray}
Using evolution equations (\ref{eq:EvEqLR}) and (\ref{eq:EvEqRR}), we derive charge continuity equations in LR and RR bases, that have the form:
\begin{eqnarray}
\frac{d}{dt}Q(x)+\partial_x J_Q(x)=\mathcal{S}.
\end{eqnarray} 
Here, $J_Q(x)$ is the current operator, and $\mathcal{S}$ is the source term. The source term physically means that there is a source or drain of the quasiparticle charge in the system. In conventional superconductors it is due to a conversion of a quasiparticle current to a condensate one \cite{blonder:prb81}.

 As the kinetic energy is Hermitian and has a quadratic spectrum, the current operator $J_Q$ is the same in LR and RR bases and has the conventional form:
\begin{eqnarray}
\label{eq:Current}
J_Q(x)=\frac{ie}{2m}\sum_{\sigma=\uparrow,\downarrow}([\partial_x\psi^\dagger_\sigma(x)]\psi_\sigma(x)-\psi^\dagger_\sigma(x)\partial_x\psi_\sigma(x)).\ \ \
\end{eqnarray}  
However, the source terms are different:
\begin{eqnarray}
&&\mathcal{S}_{LR}=e\Delta\sum_{\sigma=\uparrow,\downarrow}([\partial_x \psi_\sigma(x)]\psi_{\bar{\sigma}}(x)-\psi^\dagger_{\bar{\sigma}}(x)\partial_x\psi^\dagger_\sigma(x)),\\
\nonumber &&\mathcal{S}_{RR}=\\ \nonumber&&=e\Delta[4\int dy(\rho(x)\psi_\downarrow(y)\partial_y\psi_\uparrow(y)-\psi^\dagger_\uparrow(y)\partial_y\psi^\dagger_\downarrow(y)\rho(x))-\\&& -\sum_{\sigma=\uparrow,\downarrow}(\psi_\sigma^\dagger(x)\partial_x\psi^\dagger_{\bar{\sigma}}(x)+[\partial_x\psi_{\bar{\sigma}}(x)]\psi_{\sigma}(x))].\ \ \ \ \ 
\end{eqnarray}
Here, the terms with $\rho(x)$ in $\mathcal{S}_{RR}$ indicate exponentially growing and decaying states. $\mathcal{S}_{LR}$ and the quartic part of $\mathcal{S}_{RR}$ are Hermitian, consequently they conserve the overall charge current.

{\it Electron field operators in LR and RR representation.} In order to perform averaging, we need to find the eigenvectors and eigenvalues of $H_{\rm PTS}(k)$. For that, we express electron field operators in terms of particles and holes of Landau-Fermi liquid formalism taking into account that $H_{\rm PTS}(k)$ is non-Hermitian, i.e. Bogoliubov transformation in a non-Hermitian case. 

The Hamiltonian $H_{\rm PTS}(k)$ can be diagonalized in terms of LR basis: $D_L^\dagger H_{\rm PTS}(k)D_R={\rm diag}\{E_k,-E_k\}$. The matrices $D_L$ and $D_R$ consist of the left and right eigenvectors of $H_{\rm PTS}(k)$, respectively. For real eigenvalues $E_k$, we have:
\begin{eqnarray}
D_{L,R}=\begin{pmatrix}U^>_{R,L}(k) & U^<_{R,L}(k)\\ V^>_{R,L}(k) & V^<_{R,L}(k)\end{pmatrix}.
 \end{eqnarray}
Here, the superscript $>$ denotes eigenstates with $\Re{[E_k]}>0$ and the superscript $<$ the eigenstates with $\Re{[E_k]}<0$. For imaginary energies, we need to exchange superscripts $>$ and $<$ in $D_L$, denoting $\Im{[E]}>0$ and $\Im{[E]}<0$, respectively. Note that $D^\dagger_RH_{\rm PTS}(k)D_R$ is not diagonal. 

We express electron field operators $\Psi(k)$, $\Psi^\dagger(k)$ in terms of particles and holes in the following way: In the LR basis, $\Psi(k)=D_R\Gamma_R(k)$ and $\Psi^\dagger(k)=\Gamma_L^\dagger D_L^\dagger$. In the RR basis, we take Hermitian conjugate for $\Psi^\dagger_R(k)$: $\Psi^\dagger_R(k)=(D_R \Gamma_R(k))^\dagger$.
For clarity, we note that $\Gamma_R(k)^T=(\gamma_{R,\uparrow}(k)\ \ \gamma_{R,\downarrow}^\dagger(-k))^T$ and $\Gamma_L^\dagger(k)=(\gamma^\dagger_{L,\uparrow}(k)\ \ \gamma_{L,\downarrow}(-k))$.

\begin{figure}[tb]
	\begin{center}
		\includegraphics[width=\linewidth]{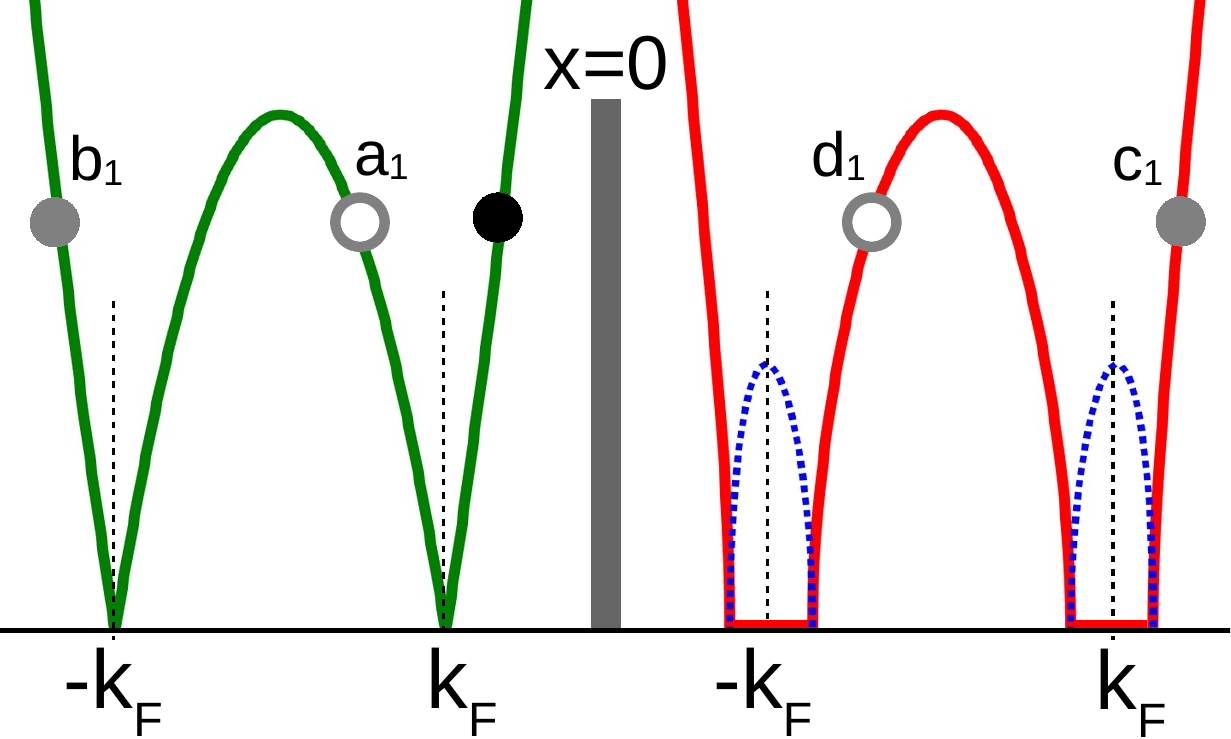}
		\caption{Scattering of a right-moving electron (black solid circle) from the normal metal at the interface with PTS through a thin insulating barrier at $x=0$. It scatters back into the normal metal as an electron (amplitude $b_1$) and, due to Andreev scattering, as a hole ($a_1$). It transfers into PTS as an electron-like quasiparticle ($c_1$) and a hole-like quasiparticle ($d_1$). The spectrum of the normal metal is shown in green. The real part of the spectrum of PTS is shown in red and the imaginary part is shown with the dotted blue line.}
		\label{fig:NIPTSelectronscattering}
	\end{center}
\end{figure}

Now, we need to study distribution functions of $\gamma$ operators. Distribution function and other thermodynamic quantities in a PT-broken regime is a very non-trivial question. Here, we have derived the distribution function in LR basis and used anticommutation relations between $\gamma$ operators \cite{supplmaterial}, obtaining $\langle\gamma_{L,\sigma}^\dagger(k_1)\gamma_{R,\sigma'}(k_2)\rangle=\langle\gamma_{R,\sigma}^\dagger(k_1)\gamma_{L,\sigma'}(k_2)\rangle=n(k_1)\delta(k_1-k_2)\delta_{\sigma,\sigma'}$ and $\langle\gamma_{R,\sigma}(k_1)\gamma^\dagger_{L,\sigma'}(k_2)\rangle=\langle\gamma_{L,\sigma}(k_1)\gamma^\dagger_{R,\sigma'}(k_2)\rangle=(1-n(k_1))\delta(k_1-k_2)\delta_{\sigma,\sigma'}$, where the distribution function is $n(k)=1/(\exp{(E_k/T)+1})$, i.e. Fermi-Dirac distribution. Notably, we obtain such distribution function assuming that creation operator acts as $\gamma_R^\dagger(k)|N_k\rangle=\sqrt{1-N_k}|1-N_k\rangle$, i.e. in a fermionic way. The Fermi-Dirac distribution was also derived in LR basis in Ref. \cite{arouca:qf22} from the partition function. However, when energy is imaginary, this function diverges, changes sign, and has an imaginary part. 

The RR basis is complete and normalized. Its wavefunctions are orthogonal with the exception of wavefunctions with the same $k$ and opposite energies. At the exceptional points $E=0$, the eigenvectors coalesce. Thus, we can define occupation numbers of the states in RR basis, if we consider only one sign of energies in our calculations of averages. This indeed holds, as we consider particle and hole excitations in Landau-Fermi liquid, where they are defined for positive energies. Thus, we will use $\langle\gamma_{R,\sigma}^\dagger(k_1)\gamma_{R,\sigma'}(k_2)\rangle=n(k_1)\delta(k_1-k_2)\delta_{\sigma,\sigma'}$. 

As the first check, we calculate the average density of electrons, $\langle\psi^\dagger_\uparrow(k)\psi_\uparrow(k)\rangle_{LR/RR}$:
\begin{eqnarray}
\label{eq:density}
\nonumber\langle\psi^\dagger_\uparrow(k)\psi_\uparrow(k)\rangle_{LR/RR}=(U_{L/R}^>(k))^*U_R^>(k)n(k)+\\ +(U_{L/R}^<(k))^*U_R^<(k)[1-n(k)].
\end{eqnarray}
In RR basis, it is $\leq 1$ with a local plateau at $E\in \Im$, where it is strictly $0.5$, because $(U_{R}^>(k))^*U_R^>(k)=(U_{R}^<(k))^*U_R^<(k)=0.5$ for $E\in \Im$. 

In LR basis, the density has divergences, which are not compensated by the wavefunctions, as it was in RR basis (for imaginary energies, the signs $>$ and $<$ should be exchanged in left wavefunctions in Eq.~(\ref{eq:density})). Since electrons in PTS are electrons in a non-equilibrium state, this is already a strong indication towards using RR basis. We have plotted the electron density in Fig. \ref{fig:ElectronDensity} for $m=1$, $\mu=10$, $T=0.01$, $\Delta k_F=1$.

\begin{figure}[tb]
	\begin{center}
		\includegraphics[width=\linewidth]{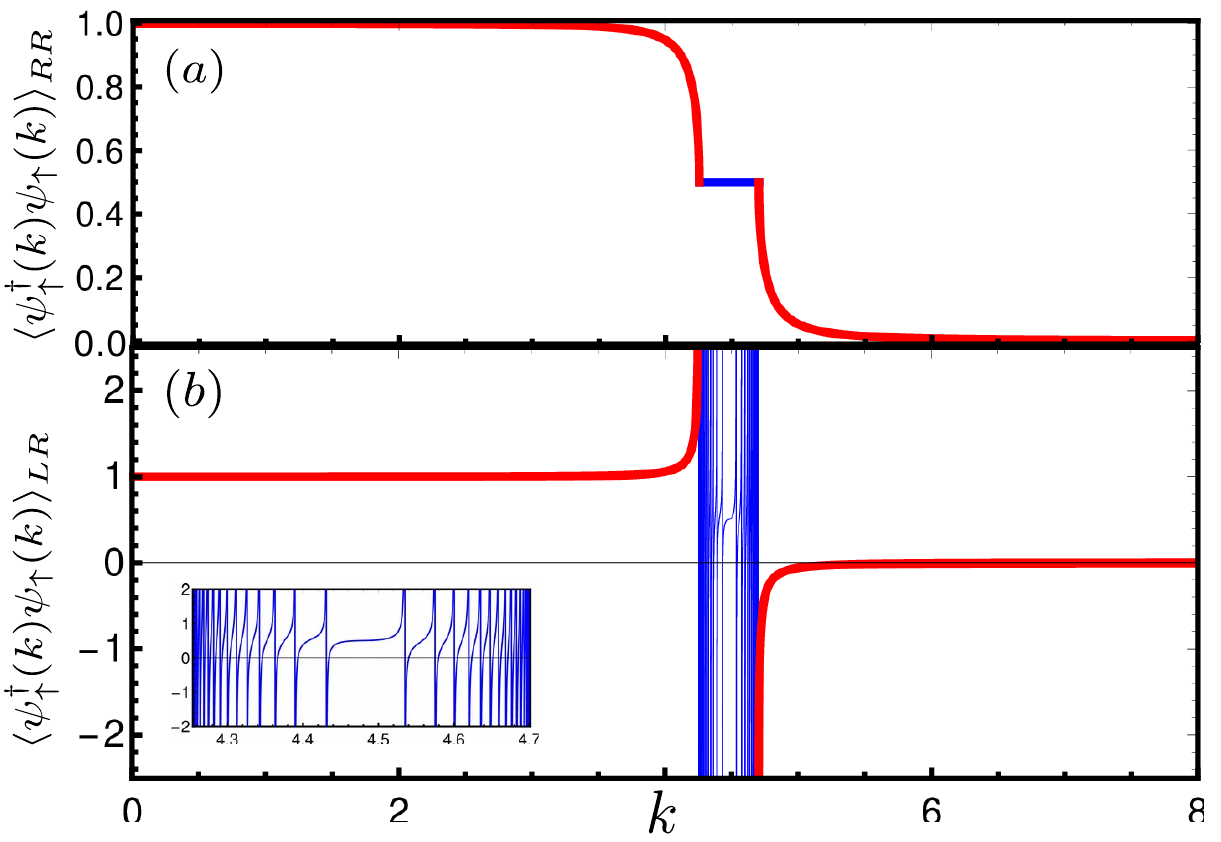}
		\caption{Averaged density of electrons with respect to momentum $k$, with red lines for momenta at real energies, and blue lines for momenta at imaginary energies (see Fig. \ref{fig:NIPTSelectronscattering}). (a) $0\leq \langle\psi^\dagger_\uparrow(k)\psi_\uparrow(k)\rangle_{RR}\leq 1$; (b) $\langle\psi^\dagger_\uparrow(k)\psi_\uparrow(k)\rangle_{LR}$ has divergences at exceptional points, $E=0$, and in the region with $E\in \Im$, see the inset.}
		\label{fig:ElectronDensity}
	\end{center}
\end{figure}

Using wavefunctions and the distribution function of $\gamma$ operators, we obtain average current $\langle J_Q\rangle_{RR/LR}$ in PTS:
 \begin{eqnarray}
\nonumber \langle J_Q\rangle_{LR/RR}=\frac{e}{m}\sum_{k}k(n(k)[(V_{L/R}^{<}(-k))^*V_R^{<}(-k)+\\ \nonumber+(U_{L/R}^{>}(k))^*U_R^{>}(k)]-(1-n(k))\times \\ 
\label{eq:currentPTS}{[(V_{L/R}^{>}(k))^*V_R^{>}(k)+(U_{L/R}^{<}(-k))^*U_R^{<}(-k)]}).\ \ \ \ \ \ \ \ 
\end{eqnarray}
For imaginary energies, we need to exchange superscripts $>$ and $<$ in left wavefunctions.

{\it Bands with imaginary energies in LR and RR bases.}
Here, we analyze bands with imaginary energies in the bulk PTS. These bands have $\Re[E]=0$, i.e. are flat bands in real energy spectrum, see Fig.~\ref{fig:NIPTSelectronscattering}. In Hermitian case, the states in flat bands have infinite density of states, but zero group velocity, i.e. they are localized. In non-Hermitian systems, the probability current is not necessarily $\propto \partial E/\partial k$ \cite{schomerus:pra14}. In PTS, the density of states is not infinite due to the imaginary part of energy. We know that $|U_{R}^{>,<}(E))|^2=|V_{R}^{>,<}(E)|^2=0.5$ for $E\in \Im$. Therefore, the current induced by any state with imaginary energy and momentum $k$ averaged in RR basis is $\langle j(k)\rangle_{RR}\propto k(2n(k)-1)$. If we integrate the quadratic part of $\mathcal{S}_{RR}$ over $x$ and average, we obtain $\langle j_{\mathcal{S}}^{RR}(k)\rangle_{RR}\propto 1-2n(k)$. Thus, the divergences in $\langle j(k)\rangle_{RR}$ and $-\langle j_{\mathcal{S}}^{RR}(k)\rangle_{RR}$ at $E\in \Im$ coincide. This looks like superconducting condensate induces huge currents acting as a whole. However, these currents are imaginary. Therefore, it is clear that we need to omit this contribution from $\langle J_Q\rangle_{RR}$. We believe that we can put $E=0$ into the distribution function, and obtain $\langle j(k)\rangle_{RR}=\langle j_{\mathcal{S}}^{RR}(k)\rangle_{RR}=0$. This means that the states are localized in these bands, because $Q(x)=e\rho(x)$.

In case of LR basis, we have: $(V_{L}^{>}(-k))^*V_R^{<}(-k)+(U_{L}^{<}(k))^*U_R^{>}(k)=1+i\Lambda=[(V_{L}^{<}(k))^*V_R^{>}(k)+(U_{L}^{>}(-k))^*U_R^{<}(-k)]^*$, where $\Lambda\in \Re$. This leads to $\langle j(k)\rangle_{LR}\propto k(2n(k)-1+i\Lambda)$, that has divergences coming from $n(E)$ at $E\in \Im$. However, the average current that is derived from $\mathcal{S}_{LR}$ is zero: $\langle j_{\mathcal{S}}^{LR}(k)\rangle_{LR}=0$. Thus, it is unclear, from where the divergences can come from. The current stays imaginary, even if we put $E=0$ in the distribution function. We will omit it, when we calculate current-voltage characteristics in LR basis below. 

We now underline the physical meaning of $\mathcal{S}_{RR}$. For $E\in \Re$, $\langle\mathcal{S}_{RR}\rangle_{RR}=0$. For $E\in \Im$, the quartic terms are in general not zero (the quadratic terms we have discussed above), where we have used the fact that Wick's theorem is valid for RR basis \cite{herviou:scipost19}. Thus, $\mathcal{S}_{RR}$ acts only in PT-broken regime and indeed denotes the growth and decay of the states with imaginary energies. Physically, this means that the non-Hermitian electron-electron interaction induces and destroys quasiparticles. Since $\mathcal{S}_{RR}$ is Hermitian, these two processes are equilibrated, i.e. as many are induced, as many are destroyed. Therefore the overall probability does not grow in the system, even though it can be not noticeable in other calculations, e.g. eigenfunctions of $H_{\rm PTS}(k)$, from where it follows that there are separate growing and decaying in time states.

{\it N-I-PTS junction.} Measuring current in PTS can be complicated due to the necessary non-equilibrium conditions applied to it. Therefore, we study the junction of a normal metal with PTS and a thin insulating barrier in between them, see Fig.~\ref{fig:NIPTSelectronscattering}. The junction is described by the BdG equations:

\begin{eqnarray}
\label{eq:BdGEq}
\nonumber\begin{pmatrix}-\frac{\partial_x^2}{2m}-\mu+I\delta(x) & -i\Delta\Theta(x)\partial_x\\ i\Delta\Theta(x)\partial_x & \frac{\partial_x^2}{2m}+\mu-I\delta(x)  \end{pmatrix}\begin{pmatrix}u(x)\\ v(x)\end{pmatrix}=\\ =E\begin{pmatrix}u(x)\\ v(x)\end{pmatrix}.\ \ 
\end{eqnarray}

Here, $I\delta(x)$ denotes insulating barrier and $\Theta(x)$ divides the normal metal part $x<0$ and the PTS part $x>0$. We search for wavefunctions in the form $\begin{pmatrix}u(x)& v(x)\end{pmatrix}=\sum_k\begin{pmatrix}u(k)& v(k)\end{pmatrix}e^{ikx}/\sqrt{L}$, where the summation is over all states that are involved in the scattering process, and $L$ is the length of the corresponding material, $L\rightarrow \infty$. In order to find these wavefunctions, we use the condition of the continuity of the wavefunction and a jump in its derivative derived from Eq. (\ref{eq:BdGEq}), see \cite{supplmaterial}. We study all scattering processes: (1) electron and (2) hole incident from the left; (3) electron-like quasiparticle and (4) hole-like quasiparticle incident from the right.

Let's consider process (1), see Fig.~\ref{fig:NIPTSelectronscattering}: an incident from the normal metal electron scatters back at the interface as an electron with the amplitude $b_1$. There is also Andreev scattering into the hole state with the amplitude $a_1$. Penetrating into the PTS particles are an electron-like quasiparticle ($c_1$) and a hole-like quasiparticle ($d_1$). All other processes are described with the amplitudes analogously, where coefficients $a$ and $d$ correspond to the reflected and transmitted particles, respectively, of the different type compared to the incident one.  

We perform this calculation for four scattering processes at positive and negative energies and for the left vectors \cite{supplmaterial}. Thus, we derive the wavefunctions $u_{R,L}^{>,<}(x)$ and $v_{R,L}^{>,<}(x)$ that describe N-I-PTS junction as a whole. We note that these wavefunctions are later used as a linear transformation, where energy substituted there has the same sign for all of them. In conventional superconductors, there are usually two types of eigenvectors: $(U\ \ V)$ and $(V^*\ \ -U^*)$ due to particle-hole symmetry. Here, the particle-hole symmetry can be defined in two ways \cite{kawabata:prx19, kornich:prr22b}, therefore we decided to explicitly take into account all eigenvectors.

The whole current through the junction is $\langle J_Q\rangle_{LR/RR}=\sum_{i=1}^4\langle j_Q^{(i)}\rangle_{LR/RR}$, where $i$ denotes scattering processes. The currents $\langle j_Q^{(i)}\rangle_{LR/RR}$ have the same formula as Eq. (\ref{eq:currentPTS}), but with $k_i$, $n_i(k_i)$, $u_{L,R}^{>,<,i}$, $v_{L,R}^{>,<,i}$ \cite{supplmaterial}. The summation is over all momenta $k_i$ that participate in the scattering process $i$. We then move to the energy representation and integrate over all necessary states. We take the density of states in PTS as $N_{\rm PTS}(E)=1/L\sum_i\delta(\Re{[E]}-\Re{[E(k_i)]})\delta(\Im{[E]}-\Im{[E(k_i)]})$, analogously to Ref. \cite{feinberg:pre99}. PTS has either real or imaginary energy, therefore $N_{\rm PTS}(E)$ eventually converts to the conventional definition of the density of states. Each $\langle j^{(i)}_Q\rangle_{LR/RR}$ contains the density of states of the incident particle in the process $i$, \cite{supplmaterial}.

{\it Comparison of average current in LR and RR bases.}
Here, we compare $\langle J_Q\rangle_{RR}$ and $\langle J_Q\rangle_{LR}$, assuming that voltage is applied to the normal metal part and is a shift of the chemical potential there \cite{supplmaterial}. In Fig. \ref{fig:currentLRRR}, we plot current-voltage characteristics of N-I-PTS junction in (a) RR and (b) LR bases. The current from the states with $E\in \Re$ is conserved through the junction, in contrast to conventional N-I-S junctions, therefore it does not matter, if we measure it in the normal metal part or PTS. We have used the parameters shown in Fig.~\ref{fig:currentLRRR} and $m=1$, $\mu=10$, $T=0.01$. Thus, the coefficients $J_0=V_0=e$.
\begin{figure}[tb]
	\begin{center}
		\includegraphics[width=0.9\linewidth]{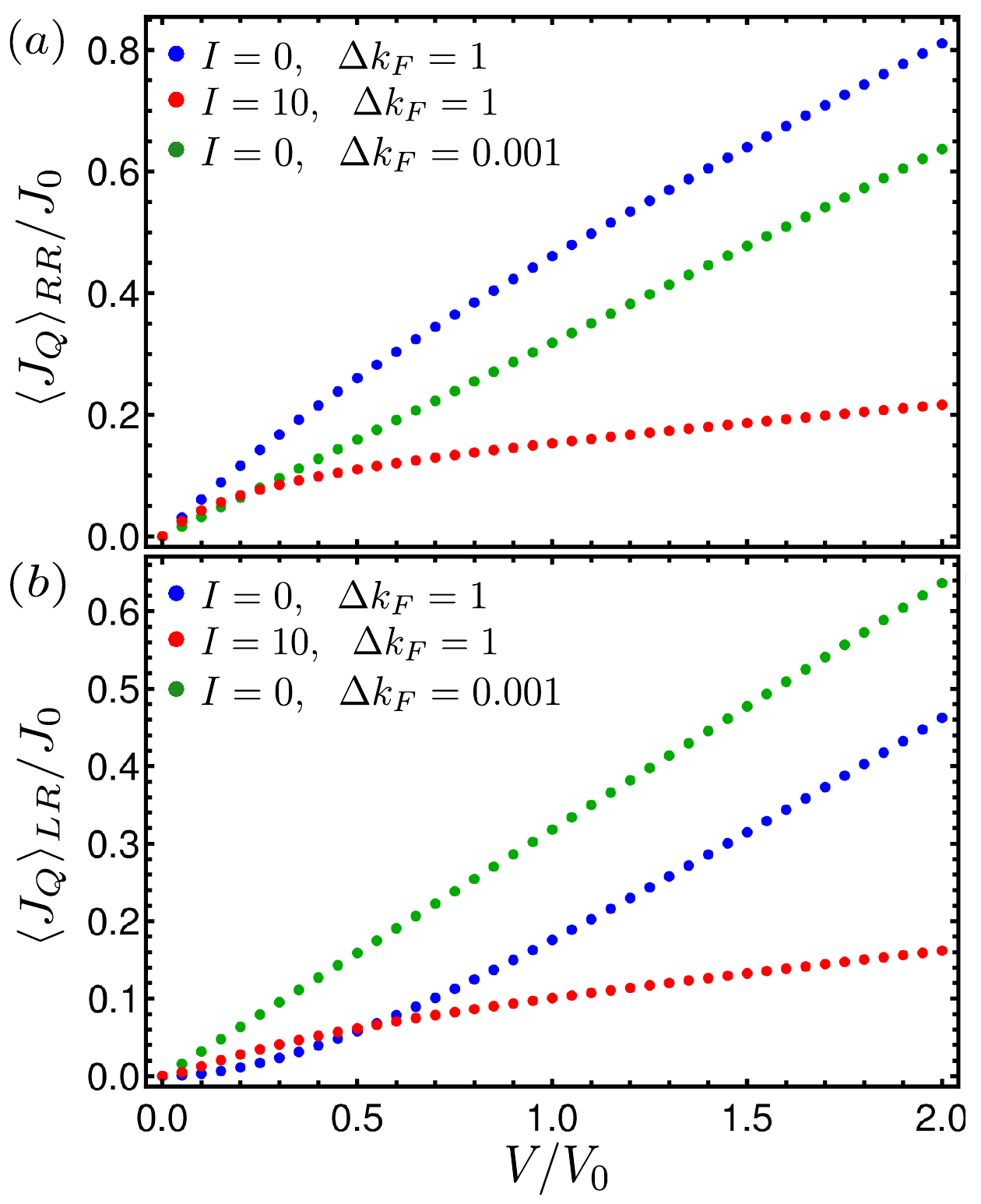}
		\caption{Current-voltage characteristics of N-I-PTS junction in (a) RR basis and (b) LR basis. The green dots correspond to no barrier ($I=0$) and almost no superconducting pairing. The blue and red dots correspond to $I=0$ and $I=10$, respectively, and a noticeable effect of superconducting pairing. Remarkably, Andreev scattering has opposite effects in different bases: (a) enhancing current; (b) suppressing current. The sign of the effect in (a) is the same as in conventional N-I-S junctions. }
		\label{fig:currentLRRR}
	\end{center}
\end{figure}

We plot the case of no barrier $I=0$ (blue and green dots) and a rather strong barrier $I=10$ (red dots). The blue and green dots have different strengths of superconducting pairing, i.e. $\Delta k_F$: the green dots represent almost N-I-N junction. The green dots show linear (Ohmic) dependence on voltage, but the blue dots in LR and RR have opposite shifts with respect to it: lower and higher, respectively. This means that the contribution from Andreev scattering has opposite signs in LR and RR bases. Indeed, in LR basis, $(a_{1,2}^L)^*a_{1,2}^R$ (for both $<$ and $>$) has the negative sign in contrast to conventional positive sign of $|a_{1,2}^R|^2$ in RR basis. For the transmitted hole and electron from the incident electron-like quasiparticle and hole-like quasiparticle, respectively, $(d_{3,4}^L)^*d_{3,4}^R$ also have negative sign for both $>$ and $<$. Thus, the conversion of electron to hole or vice versa gives the current of the opposite sign in LR basis, i.e. they move towards the junction. Scattered particles (reflected or transmitted) should move away from the scattering region unless there is some attraction force there, which is not the case here. Notably, if we change the sign of these currents, Fig.~\ref{fig:currentLRRR} (b) becomes Fig.~\ref{fig:currentLRRR} (a). 

The effect of the barrier is in general the same in LR and RR bases for large voltages: the slope is much lower for $I=10$ than for $I=0$. However, there is an opposite behaviour in LR basis for small voltages. It is due to the interplay of the current flows with $a$ and $b$ coefficients. To conclude this analysis, the exotic properties of LR basis do not eventually lead to the same observable values as RR basis, see Fig.~\ref{fig:currentLRRR}. We think that additional corrections to the definition of probability in LR formalism might eliminate the issue related to the scattering direction.

{\it Conclusions.} In this work, we show that current-voltage characteristics for the N-I-PTS junction in RR and LR bases are profoundly different, especially for the weak barrier strengths, because Andreev-scattered particles move towards the scattering region at $E\in R$ in LR basis. We argue that LR formalism is not universal and does not apply to this setup taking into account the unphysical Andreev scattering, the divergences in the density of electrons, and imaginary current contributions from the bands with $E\in \Im$. Importantly, we have shown that there is no infinite probability growth in RR basis, but the growth and decay are equilibrated, in contrast to results usually obtained for PT-symmetric non-Hermitian systems.
\begin{acknowledgments}
	I acknowledge useful discussions with Bj\"orn Trauzettel, Zhuo-Yu Xian, Jens Bardarson, and Teun Klapwijk. This work was supported by the W\"urzburg-Dresden Cluster of Excellence on Complexity and Topology in Quantum Matter (EXC2147, project-id 390858490) and by the DFG (SFB1170 ``ToCoTronics''). I thank the Bavarian Ministry of Economic Affairs, Regional Development and Energy for financial support within the High-Tech Agenda Project ``Bausteine f\"ur das Quanten Computing auf Basis topologischer Materialen''.
\end{acknowledgments}

\begin{widetext}
\section*{Supplemental Material }

\maketitle
\renewcommand{\theequation}{S\arabic{equation}}
\setcounter{equation}{0}
\renewcommand{\thefigure}{S\arabic{figure}}
\renewcommand{\figurename}{Supplementary Fig.}

\setcounter{figure}{0}
\renewcommand{\thesection}{S\arabic{section}}
\setcounter{section}{0}
In this Supplemental Material, we present additional details and calculations regarding: 1) derivation of the evolution equations in LR and RR bases; 2) derivation of the distribution functions in LR and RR bases; 3) derivation of the  wavefunctions in N-I-PTS junction; 4) calculation of the average current. 

\section{S1. Derivation of evolution equations in LR and RR bases}
By definition, right and left states are
\begin{eqnarray}
i\frac{\partial}{\partial t}|R\rangle=H|R\rangle,\\
i\frac{\partial}{\partial t}|L\rangle=H^\dagger|L\rangle.
\end{eqnarray}
Thus, their time evolution is
\begin{eqnarray}
|R(t)\rangle=e^{-iHt}|R(0)\rangle,\\
|L(t)\rangle=e^{-iH^\dagger t}|L(0)\rangle.
\end{eqnarray} 
Let's consider operator $O$ that does not depend on time in Schr\"odinger representation. An average of an operator $O$ in LR and RR bases are
\begin{eqnarray}
\label{eq:evolLR}
\langle L|O|R\rangle&=&\langle L(0)|e^{iHt}Oe^{-iHt}|R(0)\rangle,\\
\label{eq:evolRR}
\langle R|O|R\rangle&=&\langle R(0)|e^{iH^\dagger t}Oe^{-iHt}|R(0)\rangle=\langle R(0)|O^H|R(0)\rangle,
\end{eqnarray}
respectively.
We can see that Eq. (\ref{eq:evolLR}) leads to a Heisenberg evolution equation (see Eq. (3) in the main text). Eq. (\ref{eq:evolRR}), however, leads to
\begin{eqnarray}
\frac{d}{dt}O^H=i(H^\dagger O^H-O^H H),
\end{eqnarray}
i.e. Eq. (4) from the main text.

\section{S2. Derivation of Fermi-Dirac distribution function}
Let's first consider LR basis. We define the states according to their occupation number:
\begin{eqnarray}
|R\rangle=|n_0,n_1,n_2,n_3,...,n_i,...\rangle=\prod_{k=0}^\infty (\gamma^\dagger_R(k))^{n_k}|0\rangle,\\
\langle L|=\langle n_0,n_1,n_2,n_3,...,n_i,..|=\prod_{k=0}^\infty \langle 0|(\gamma_L(k))^{n_k}.
\end{eqnarray}
The averaging is defined as 
$\langle...\rangle=\frac{{\rm Tr}[...\rho]}{{\rm Tr} \rho}$
where $\rho$ is the density matrix, $\rho=e^{-\beta H}$. In the following, we modify  the conventional derivation of Fermi-Dirac distribution function (see e.g. Ref. \onlinecite{reif:book85}) according to LR basis formalism. 

Let's first average $\gamma_{L,\downarrow}(k_1)\gamma_{R,\downarrow}^\dagger(k_2)$:
\begin{eqnarray}
\langle \gamma_{L,\downarrow}(k_1)\gamma_{R,\downarrow}^\dagger(k_2)\rangle=\frac{\sum_m\langle L_m|\gamma_{L,\downarrow}(k_1)\gamma_{R,\downarrow}^\dagger(k_2)e^{-\beta H}|R_m\rangle}{\sum_{n_1',n_2',n_3',..}e^{-\beta \sum_NE_Nn_N'}}=\\ \nonumber=\frac{\sum_m\langle L_m|\gamma_{L,\downarrow}(-k_1)\gamma_{R,\downarrow}^\dagger(-k_2)|R_m\rangle e^{-\beta\sum_n E_{n} n_{m_n}}}{\sum_{n_1',n_2',n_3',..}e^{-\beta \sum_NE_Nn_N'}}=\frac{\sum_m (1-n_{m_{k_1}})\delta_{k_1,k_2}e^{-\beta \sum_nE_nn_{m_n}}}{\sum_{n_1',n_2',n_3',..}e^{-\beta \sum_NE_Nn_N'}}=\\ \nonumber=\delta_{k_1,k_2}-\delta_{k_1,k_2}\frac{\sum_m n_{m_{k_1}}e^{-\beta \sum_nE_nn_{m_n}}}{\sum_{n_1',n_2',n_3',..}e^{-\beta \sum_NE_Nn_N'}}=\delta_{k_1,k_2}-\delta_{k_1,k_2}\frac{\sum_{n_1,n_2,n_3...} n_{k_1}e^{-\beta \sum_iE_in_{i}}}{\sum_{n_1',n_2',n_3',..}e^{-\beta \sum_NE_Nn_N'}}=\\ \nonumber=\delta_{k_1,k_2}-\delta_{k_1,k_2}\frac{\sum_{n_{k_1}}n_{k_1}e^{-\beta E_{k_1}n_{k_1}}\sum_{n_1,n_2,n_3,..n_{k_1-1},n_{k_1+1},..} e^{-\beta(E_1n_1+E_2n_2+E_3n_3+...)}}{\sum_{n'_{k_1}}e^{-\beta E_{k_1}n'_{k_1}}\sum_{n_1',n_2',n_3',.., n'_{k_1-1},n'_{k_1+1},..}e^{-\beta(E_1n'_1+E_2n'_2+E_3n'_3+...)}}.
\end{eqnarray}
Let's denote the sum without $n_{k_1}$ as $Z_{n_k}(N-n_{k_1})=\sum_{n_1,n_2,n_3,.., n_{k_1-1},n_{k_1+1},..}e^{-\beta(E_1n_1+E_2n_2+E_3n_3+...)}$, where $N$ is the number of all particles. Note, that it still depends on $n_{k_1}$, as the number of particles is conserved. We assume that the following approximation is valid:
\begin{eqnarray}
\ln{Z_{k_1}(N-1)}\simeq \ln Z_{k_1}(N)-\frac{\partial \ln{Z_{k_1}(N)}}{\partial N}=\ln{Z_{k_1}(N)}-\alpha_{k_1}.
\end{eqnarray}
If $N$ is large enough that the chemical potential does not change significantly, when a particle is added or subtracted, and taking into account that free energy $F=-\ln{Z}/\beta$, we obtain
\begin{eqnarray} 
\alpha_{k_1}=-\beta\frac{\partial F}{\partial N}= -\mu\beta.
\end{eqnarray}
Then, we obtain
\begin{eqnarray}
\frac{Z_{k_1}(N)}{Z_{k_1}(N-1)}=e^{-\mu\beta}.
\end{eqnarray}
Altogether, we obtain
\begin{eqnarray}
\langle \gamma_{L,\downarrow}(k_1)\gamma_{R,\downarrow}^\dagger(k_2)\rangle=\delta_{k_1,k_2}-\delta_{k_1,k_2}\frac{0+1e^{-\beta E_{k_1}}Z_{k_1}(N-1)}{Z_{k_1}(N)+e^{-\beta E_{k_1}}Z_{k_1}(N-1)}=\delta_{k_1,k_2}-\frac{\delta_{k_1,k_2}}{e^{\beta E_{k_1}}\frac{Z_{k_1}(N)}{Z_{k_1}(N-1)}+1}=\\ \nonumber=\delta_{k_1,k_2}\left(1-\frac{1}{e^{(E_{k_1}-\mu)\beta}+1}\right).
\end{eqnarray}
The operators $\gamma_L^\dagger(k)$ and $\gamma_R(k)$ destroy the state $k$ in the left and right states, respectively. Thus, the average $\langle\gamma^\dagger_{L,\uparrow}(k_1)\gamma_{R,\uparrow}(k_2)\rangle=\delta(k_1-k_2)n_{k_1}$. 

From the relation between $\gamma$ and $\psi$ operators follows that the anticommutation relations between left and right operators $\gamma$ are:
\begin{eqnarray}
\{\gamma_{R,\uparrow}(k_1),\gamma^\dagger_{L,\uparrow}(k_2)\}=\delta(k_1-k_2),\\
\{\gamma_{R,\downarrow}^\dagger(k_1), \gamma_{L,\downarrow}(k_2)\}=\delta(k_1-k_2).
\end{eqnarray}
Therefore, in order to obtain two other averages, we exchange left and right operators, obtaining
\begin{eqnarray}
\langle\gamma_{R,\uparrow}(k_1)\gamma_{L,\uparrow}^\dagger(k_2)\rangle&=&\delta(k_1-k_2)(1-n_{k_1}),\\
\langle\gamma^\dagger_{R,\downarrow}(k_1)\gamma_{L,\downarrow}(k_2)\rangle&=&\delta(k_1-k_2)n_{k_1}.
\end{eqnarray}

In case of RR basis, the states are orthogonal, except for the states with the same momentum $k$ and different signs of energy. This means that we can use the same derivation of a distribution function as above, provided we consider energies of only one sign. At $E=0$ we have exceptional points, where eigenvectors coalesce. Taking into account that operators $\gamma$ represent particle and hole excitations in Landau-Fermi liquid, that are defined at positive energy, this constraint holds.

\section{S3. Derivation of the wavefunctions in N-I-PTS junction}
The BdG equations for the quasiparticle wavefunctions in a N-I-PTS junction are
\begin{eqnarray}
\label{eq:HNPTS}
\begin{pmatrix}-\frac{\partial_x^2}{2m}-\mu+I\delta(x) & -i\Delta\Theta(x)\partial_x\\ i\Delta\Theta(x)\partial_x & \frac{\partial_x^2}{2m}+\mu-I\delta(x)  \end{pmatrix}\begin{pmatrix}u(x)\\ v(x)\end{pmatrix}=E\begin{pmatrix}u(x)\\ v(x)\end{pmatrix}.
\end{eqnarray}
There is a delta-function barrier between N and PTS with the strength $I$ at $x=0$. This means, the wavefunction is continuous at $x=0$:
\begin{eqnarray}
\label{eq:continuity}
\begin{pmatrix}u(x)\\ v(x)\end{pmatrix}\Bigg|_{x=-0}=\begin{pmatrix}u(x)\\ v(x)\end{pmatrix}\Bigg|_{x=+0},\end{eqnarray} 
and there is a step in the derivative of the wavefunction, that appears when integrating left side of Eq. (\ref{eq:HNPTS}) from $-0$ to $+0$:
\begin{eqnarray}
\label{eq:derivative}
\partial_x\begin{pmatrix}u(x)\\ v(x)\end{pmatrix}\Bigg|_{x=+0}-\partial_x\begin{pmatrix}u(x)\\ v(x)\end{pmatrix}\Bigg|_{x=-0}=2mI\begin{pmatrix}u(0)\\ v(0)\end{pmatrix}.
\end{eqnarray}
We search for the wavefunctions of quasiparticles in N-I-PTS junction of the form of
\begin{eqnarray}
\begin{pmatrix}u(x)\\ v(x)\end{pmatrix}=\begin{pmatrix}u(k)\\v(k)\end{pmatrix}e^{ikx}.
\end{eqnarray}
Applying it to Eq. (\ref{eq:HNPTS}) separately for $x>0$ and $x<0$, we obtain the wavevectors: $K_{1,2}=\sqrt{2m(\mu\pm E)}$ for the normal metal and 
\begin{eqnarray}
\label{eq:Rootsk}
k_{1,2}=\pm\sqrt{2m}\sqrt{\Delta^2m+\mu\pm\sqrt{E^2+m^2\Delta^4+2\Delta^2m\mu}},
\end{eqnarray}
for PTS.

There are four possible scattering processes: 1) electron incident on the junction from the left; 2) hole incident on the junction from the left; 3) electron-type quasiparticle incident on the junction from the right; 4) hole-type quasiparticle incident on the junction from the right. The wavefunctions for them for scattering at $E>0$ with the corresponding indices are:
\begin{eqnarray}
&&\begin{pmatrix}1\\0\end{pmatrix} e^{i K_1 x}+a^>_1 \begin{pmatrix}0 \\1\end{pmatrix}e^{i K_2 x}+b^>_1\begin{pmatrix}1\\ 0\end{pmatrix}e^{-i K_1 x}\Bigg|_{x=-0}=c^>_1\begin{pmatrix}U^>(k_1)\\ V^>(k_1)\end{pmatrix}e^{i k_1x}+d^>_1\begin{pmatrix}U^>(-k_2)\\ V^>(-k_2)\end{pmatrix}e^{-i k_2 x}\Bigg|_{x=+0},\\
&&\begin{pmatrix}0\\ 1\end{pmatrix}e^{-iK_2x}+a_2^>\begin{pmatrix}1\\ 0\end{pmatrix}e^{-iK_1x}+b_2^>\begin{pmatrix}0\\ 1\end{pmatrix}e^{iK_2x}\Bigg|_{x=-0}=d_2^>\begin{pmatrix}U^>(k_1)\\ V^>(k_1)\end{pmatrix}e^{i k_1x}+c_2^>\begin{pmatrix}U^>(-k_2)\\ V^>(-k_2)\end{pmatrix}e^{-i k_2x}\Bigg|_{x=+0},\\
&&\begin{pmatrix}U^>(-k_1)\\ V^>(-k_1)\end{pmatrix}e^{-ik_1x}+a_3^>\begin{pmatrix}U^>(-k_2)\\ V^>(-k_2)\end{pmatrix}e^{-ik_2x}+b_3^>\begin{pmatrix}U^>(k_1)\\ V^>(k_1)\end{pmatrix}e^{ik_1x}\Bigg|_{x=+0}=c_3^>\begin{pmatrix}1\\0\end{pmatrix}e^{-iK_1x}+d_3^>\begin{pmatrix}0\\1\end{pmatrix}e^{iK_2x}\Bigg|_{x=-0}, \ \ \ \  \\
&&\begin{pmatrix}U^>(k_2)\\V^>(k_2)\end{pmatrix}e^{ik_2x}+a_4^>\begin{pmatrix}U^>(k_1)\\ V^>(k_1)\end{pmatrix}e^{ik_1x}+b_4^>\begin{pmatrix}U^>(-k_2)\\ V^>(-k_2)\end{pmatrix}e^{-ik_2x}\Bigg|_{x=+0}=c_4^>\begin{pmatrix}0\\1\end{pmatrix}e^{iK_2x}+d_4^>\begin{pmatrix}1\\0\end{pmatrix}e^{-iK_1x}\Bigg|_{x=-0}.
\end{eqnarray}
For scattering at $E<0$, we have
\begin{eqnarray}
&&\begin{pmatrix}1\\0\end{pmatrix} e^{i K_2 x}+a_1^< \begin{pmatrix}0 \\1\end{pmatrix}e^{i K_1 x}+b_1^<\begin{pmatrix}1\\ 0\end{pmatrix}e^{-i K_2 x}\Bigg|_{x=-0}=c_1^<\begin{pmatrix}U^<(k_2)\\ V^<(k_2)\end{pmatrix}e^{i k_2x}+d_1^<\begin{pmatrix}U^<(-k_1)\\ V^<(-k_1)\end{pmatrix}e^{-i k_1 x}\Bigg|_{x=+0},\\
&&\begin{pmatrix}0\\ 1\end{pmatrix}e^{-iK_1x}+a_2^<\begin{pmatrix}1\\ 0\end{pmatrix}e^{-iK_2x}+b_2^<\begin{pmatrix}0\\ 1\end{pmatrix}e^{iK_1x}\Bigg|_{x=-0}=d_2^<\begin{pmatrix}U^<(k_2)\\ V^<(k_2)\end{pmatrix}e^{i k_2x}+c_2^<\begin{pmatrix}U^<(-k_1)\\ V^<(-k_1)\end{pmatrix}e^{-i k_1x}\Bigg|_{x=+0},\\
&&\begin{pmatrix}U^<(-k_2)\\ V^<(-k_2)\end{pmatrix}e^{-ik_2x}+a_3^<\begin{pmatrix}U^<(-k_1)\\ V^<(-k_1)\end{pmatrix}e^{-ik_1x}+b_3^<\begin{pmatrix}U^<(k_2)\\ V^<(k_2)\end{pmatrix}e^{ik_2x}\Bigg|_{x=+0}=c_3^<\begin{pmatrix}1\\0\end{pmatrix}e^{-iK_2x}+d_3^<\begin{pmatrix}0\\1\end{pmatrix}e^{iK_1x}\Bigg|_{x=-0}, \ \ \ \ \ \\
&&\begin{pmatrix}U^<(k_1)\\V^<(k_1)\end{pmatrix}e^{ik_1x}+a_4^<\begin{pmatrix}U^<(k_2)\\ V^<(k_2)\end{pmatrix}e^{ik_2x}+b_4^<\begin{pmatrix}U^<(-k_1)\\ V^<(-k_1)\end{pmatrix}e^{-ik_1x}\Bigg|_{x=+0}=c_4^<\begin{pmatrix}0\\1\end{pmatrix}e^{iK_1x}+d_4^<\begin{pmatrix}1\\0\end{pmatrix}e^{-iK_2x}\Bigg|_{x=-0}.
\end{eqnarray}
We then find coefficients $a_{1,2,3,4}^{>,<}$, $b_{1,2,3,4}^{>,<}$, $c_{1,2,3,4}^{>,<}$, and $d_{1,2,3,4}^{>,<}$ from Eqs. (\ref{eq:derivative}) and (\ref{eq:continuity}). We do this derivation for the left and right wavefunctions.

\section{S4. Calculation of the average current}
Let's assume that the current is measured in the normal metal. Then, after averaging and expressing the sum over scattering states in energy representation, the contribution to the current from the process 1) is:
\begin{eqnarray}
\langle j_Q^{(1)}\rangle_{LR}=\frac{e}{m}\int dE(N_n(K_1)[K_1n_1(E)-K_1\left(b_1^{>,L}\right)^*b_1^{>,R}n_1(E) -K_2\left(a_1^{>,L}\right)^*a_1^{>,R}(1-n_1(E))]+\\ \nonumber N_n(K_2)[n_1(E)K_1\left(a_1^{<,L}\right)^*a_1^{<,R}-(1-n_1(E))K_2+(1-n_1(E))K_2\left(b_1^{<,L}\right)^*b_1^{<,R}]).
\end{eqnarray}
Here, $N_n$ is the density of states in the normal metal. For the other three processes, we have
\begin{eqnarray}
&&\langle j_Q^{(2)}\rangle_{LR}=\frac{e}{m}\int dE(N_n(K_2)[-K_1(a_2^{>,L})^*a_2^{>,R}n_2(E)+K_2(1-n_2(E))-K_2(b_2^{>,L})^*b_2^{>,R}(1-n_2(E))]+\\ \nonumber &&+N_n(-K_1)[(-K_1)n_2(E)+K_1n_2(E)(b_2^{<,L})^*b_2^{<,R}-(1-n_2(E))(-K_2)(a_2^{<,L})^*a_2^{<,R}]),\\
&&\langle j_Q^{(3)}\rangle_{LR}=\frac{e}{m}\int dE(N_{\rm PTS}(-k_1)[-K_1(c_3^{>,L})^*c_3^{>,R}n_3(E)-K_2(d_3^{>,L})^*d_3^{>,R}(1-n_3(E))]+\\ \nonumber&& +N_{\rm PTS}(-k_2)[n_3(E)K_1(d_3^{<,L})^*d_3^{<,R}-(1-n_3(E))(-K_2)(c_3^{<,L})^*c_3^{<,R}]),\\
&&\langle j_Q^{(4)}\rangle_{LR}=\frac{e}{m}\int dE(N_{\rm PTS}(k_2)[-K_1(d_4^{>,L})^*d_4^{>,R}n_4(E)-K_2(c_4^{>,L})^*c_4^{>,R}(1-n_4(E))]+\\ \nonumber&& +N_{\rm PTS}(k_1)[n_4(E)K_1(c_4^{<,L})^*c_4^{<,R}-(1-n_4(E))(-K_2)(d_4^{<,L})^*d_4^{<,R}]).
\end{eqnarray}
The whole current is $\langle J_Q\rangle_{LR}=\sum_{i=1}^4\langle j_Q^{(i)}\rangle_{LR}$. In order to obtain $\langle J_Q\rangle_{RR}$, we need to substitute the left wavefunctions with the right ones.

As we assume that voltage is applied to the normal metal part of the junction, the distribution function for the process 1) is $n_1(E)=n(E-V)=1/(\exp{[(E-V)/T]}+1)$. For the process 2), we have $n_2(E)=1-n(-E-V)$, because the hole excitation has the same distribution function as the absence of an electron at the energy $-E$: $1-n(-E)\equiv n(E)$. Therefore, due to the different physical meaning of the hole excitations, voltage should be introduced into the distribution function $n_2(E)$ differently than in $n_1(E)$. For processes 3) and 4), $n_3(E)=n_4(E)=n(E)$, but analogously with the different physical meaning.

\end{widetext}

\end{document}